# Studying a Technical scheme for Accurate Measurement of Gravity Waves Momentum[†]

Farzin Naserian,[1, *]

[1] Faculty of Physics, Shahid Bahonar University of Kerman, Kerman, Iran

**abstract**

In spite of all the attempts conducted to improve the accuracy of the gravity wave detectors in recent years, no method has been successful to measure these waves up to now. Most of these detectors and laser interferometers work based on the position measurement, because the use of such kind of measurement leads to decreasing the accuracy of measurement, since it contributes to main problems such as requiring strong interaction among different components of the system. Maybe the main cause of failing is applying this kind of measurement.

A method in improving the accuracy of these detectors and decreasing such problems is applying the momentum measurement. At present, no practical way to measure momentum, in this paper, a theoretical scheme is presented to measure momentum.

## I. Introduction

The precise definition by Kip Thorne, `gravitational waves are ripples in the curvature of space-time that are emitted by violent astrophysical events, and that propagate out from their source with the speed of light' [1,2]. A weak GW far away from its birthplace can be most easily understood from analyzing its action on the probe bodies motion in some region of space-time.

The fast progress in improving the sensitivity of the gravitational-wave (GW) detectors, we all have witnessed in the recent years, has propelled the scientific community to the point, when quantum behavior of such immense measurement devices as kilometer-long interferometers starts to matter.

The more-than-ten-years-long history of the large-scale laser gravitation-wave detectors can be considered both as a great success and a complete failure, depending on the point of view. On the one hand, virtually all technical requirements for these detectors have been met, and the planned sensitivity levels have been achieved. On the other hand, no gravitational waves have been detected thus far.

Gravity wave detectors consist of aluminum (or sapphire or silicon or niobium) bars, weighing between 10 kilograms and 10 tons, which are driven into motion by passing waves of gravity. The motions are very tiny: for the gravity waves that theorists predict are bathing the earth, a displacement $\delta x \cong 10^{-19}$ might be typical. And this displacement may oscillate, due to oscillations of the gravity wave, with a period $p \cong 10^{-3}$ second.

An "initial" measurement of the bar's east-west position with precision $\Delta x_i \cong 10^{-19}$cm will inevitably disturb the bar's east west momentum by $\Delta p \geq \frac{\hbar}{2\Delta x_i}$, and correspondingly will disturb its velocity by $\Delta v = \frac{\Delta p}{m} \geq \frac{\hbar}{2m\Delta x_i}$, where m is bar's mass. During the time interval $\tilde{\tau} \sim 10^{-3}$ second between measurements, the mass will move away from its initial position by an amount, $\Delta x_m = \Delta v \tilde{\tau} \geq \frac{\hbar \tilde{\tau}}{2m\Delta x_i}$, which is unpredictable because $\Delta v$ is unpredictable. Putting in numbers ($\tilde{\tau} = 10^{-3}$s ، m = 10 tons ،$\Delta x_i \cong 10^{-19}$cm), we find $\Delta x_m \geq 5 \times 10^{-19}$cm-which is some what larger than the desired precision of our sequence of measurements. If the next measurement reveal a position changed by as much as $5 \times 10^{-19}$cm, we have no way of knowing whether the change was due to a passing gravity wave or to the unpredictable, quantum mechanical disturbance made by our first measurement plus subsequent free motion of the bar has "demolished" all possibility of making a second measurement of the same precision, $\Delta x \sim 10^{-19}$cm, as the first, and of thereby monitoring the bar and detecting the expected gravity waves.

In principle one can circumvent this problem by making the bar much heaver than 10 tons (recall that $\Delta x_m$ is inversely proportional to the mass). However, this is impractical. In principle another solution is to shorten the time between measurements ( recall that $\Delta x_m$ is directly proportional to $\tilde{\tau}$). However this will weaken the gravitational-wave signal ($\delta x_{GW} \propto \tilde{\tau}^2$ for $\tilde{\tau} \leq 10^{-3}$ second) even more than it reduces the unpredictable movement of the bar($\Delta x_m \propto \tilde{\tau}$).

The best solution is cleverness: find some way to make the gravity-wave effect stronger; this is being done in laser interferometer gravity-wave detectors [3], but only at the price of having to make $10^{-16}$cm measurements of the relative displacement of two bars as far apart as several kilometers. Alternatively, find some way to circumvent the effects of the Heisenberg uncertainty principle- that is, some way to prevent the inevitable disturbance due to the first measurement, plus subsequent free motion, from demolishing the possibility of a second accurate measurement: a quantum non-demolition (QND) method.

One QND method which could work in principle is this: instead of measuring the position of the 10-ton bar, measure its momentum with a small enough initial error, $\Delta p_i \sim 10^{-9} \, \text{gcm}/\text{sec}$, to detect the expected gravity waves. Thereby inevitably the bar's position by an unknown amount $\Delta x \geq \frac{\hbar}{2\Delta p_i} \sim 5 \times 10^{-19} \, \text{cm}$. Wait a time $\tilde{\tau} \sim 10^{-3}$ second and then make another momentum measurement. As the bar moves freely between the measurements, its momentum remains fixed. The uncertainty $\Delta x$ in the bar's position does not by free evolution produce a new uncertainty $\Delta p_m$ in the momentum. Consequently the second measurement can have as good accuracy, $10^{-9} \, \text{gcm}/\text{sec}$, as the initial measurement: and a momentum change of (a few) $\times 10^{-9} \, \text{gcm}/\text{sec}$ due to a passing gravity wave can be seen.

Momentum measurements can be quantum non-demolition, but position measurements cannot be, for this simple reason: in its free motion between measurements the bar keeps its momentum constant, but it changes its position by an amount $\delta x = \left(\frac{p}{m}\right)\tilde{\tau}$ that depends on the momentum, and that therefore is uncertain because of measurement-induced uncertainties of the momentum. We say that momentum is a QND variable, but position is not.

Unfortunately, however, it is far easier to measure position than momentum. Up to now no one has discovered a realizable method in technical sense for momentum measuring with wanted accuracy but in this paper, a theoretical technical method for momentum measuring will be studied.

A resonant-bar gravity-wave antenna is an oscillator with mass m, frequency ω, position $\hat{x}$, and momentum $\hat{p}$, which couples to a gravitational waves (classical external force F) with a coupling energy $\hat{H}_F = -\hat{x}F(t)$. In most experiments the antenna's position $\hat{x}$ is coupled by a transducer ($\hat{H}_I = K\hat{x}\hat{q}$ ; $K \equiv$ coupling constant) to an electromagnetic circuit (quantum readout system), which we shall describe as an oscillator with capacitance C, inductance L, generalized coordinate (equal to charge on the capacitor) $\hat{q}$, and generalized momentum (equal to flux inductor) $\hat{\pi}$. More complicated QRS's can be used; but this is the typical case. The voltage on the capacitor, which is proportional to $\hat{q}$, is monitored by an amplifier-the first classical stage of the measuring system. Thus $\hat{q}$ is the readout observable $\hat{Q}_R$.

The coupled antenna, force, and QRS are governed by the Hamiltonian

$$\hat{H} = \frac{\hat{p}^2}{2m} + \frac{1}{2}m\omega^2\hat{x}^2 + \frac{\hat{\pi}^2}{2L} + \frac{\hat{q}^2}{2C} + \hat{H}_F + \hat{H}_l$$

$$\hat{H}_F = -\hat{x}F(t) \quad , \quad \hat{H}_l = K\hat{x}\hat{q} \tag{1}$$

For which the Heisenberg evolution equations are

$$\frac{d\hat{x}}{dt} = \frac{\hat{p}}{m} \;,\; \frac{d\hat{p}}{dt} = -m\omega^2\hat{x} + F(t) - K\hat{q}$$

$$\frac{d\hat{q}}{dt} = \frac{\hat{\pi}}{L} \;,\; \frac{d\hat{\pi}}{dt} = \frac{-\hat{q}}{C} - K\hat{x} \tag{2}$$

Because these equations ignore the first classical stage (amplifier) and its detailed back action on the QRS, they cannot tell us the actual sensitivity of the measuring system. On the other hand, they can tell us the ultimate quantum mechanical limit on the sensitivity.

Suppose, as a first case, that the signal $\hat{Q}_R = \hat{q}$ is fed continuously into the amplifier for a time much longer than a quarter-cycle of the antenna, and that one's goal is to measure $\hat{x}_0$.

For an oscillator the conserved quantities, which are guaranteed to be QND observables at any and all times, include the energy [4] and the real and imaginary parts of the complex amplitude [5]:

$$\hat{X}_1 = \hat{x}(t)\cos\omega t - \left(\frac{\hat{p}(t)}{m\omega}\right)\sin\omega t \tag{3}$$

$$\hat{X}_2 = \hat{x}(t)\sin\omega t - \left(\frac{\hat{p}(t)}{m\omega}\right)\cos\omega t \tag{4}$$

High-precision measurements of $\hat{X}_1$ or $\hat{X}_2$ (whether fully QND or not) are called back-action-evolution measurements [6,7], because they enable the measured component of the amplitude (for example $\hat{X}_1$) to avoid back-action contamination by the measuring device, at the price of strongly contaminating the other component ($\hat{X}_2$). (The uncertainty relation $\Delta\hat{X}_1\Delta\hat{X}_2 \geq \frac{\hbar}{2m\omega}$ (5) is enforced by the commutation relations $[\hat{X}_1, \hat{X}_2] = \frac{i\hbar}{m\omega}$, which follow from $[\hat{x}, \hat{p}] = i\hbar$.)

During the measurement $\hat{x}(t)$, which feeds $\hat{\pi}$ and thence $\widehat{Q}_R \equiv \hat{q}$, oscillates between $\hat{x}_0$ and $\hat{p}_0$. [$\hat{x}(t) = \hat{x}_0 \cos \omega t + \left(\frac{\hat{p}_0}{m\omega}\right) \sin \omega t$, aside from minor modifications due to the coupling. Note that $\hat{x}_0 \equiv \widehat{X}_1$, $\frac{\hat{p}_0}{m\omega} \equiv \widehat{X}_2$; Eqs. 3,4,5.] Consequently, the signal $\widehat{Q}_R$ entering the amplifier contains not only $\hat{x}_0$ but also, unavoidably, $\hat{p}_0$. Since their relative strengths in the signal are $\frac{p_0}{x_0} = m\omega$, the measurement determines them with relative precisions $\Delta p_0 = m\omega \Delta x_0$. Taking account of the uncertainty relation $\Delta x_0 \Delta p_0 \geq \frac{\hbar}{2}$, we find [6,8,9,10]:

$$\Delta x_0 = \frac{\Delta p_0}{m\omega} \geq \left(\frac{\hbar}{2m\omega}\right)^{1/2} \qquad (6).$$

Such a measurement is called an amplitude-and-phase measurement because it gives information about both the amplitude $\left[x_0^2 + \left(\frac{p_0}{m\omega}\right)^2\right]^{1/2}$ and the phase $\psi_0 = \tan^{-1}\left(\frac{p_0}{m\omega x_0}\right)$ of the antenna's motions. An ideal amplitude-and-phase measurement with the limiting sensitivity in Eq.6 drives the antenna into a coherent (minimal-wave-packet) state. If such a measurement (state preparation) has put the antenna into a coherent state with $\langle \hat{x}(t) \rangle = x_0 \cos \omega t + \left(\frac{p_0}{m\omega}\right) \sin \omega t$, then a classical force $F = F_0 \cos(\omega t + \varphi)$ acting for a time $\tilde{\tau} \gg \frac{2\pi}{\omega}$ will leave the state coherent but changes its amplitudes by $\delta x_0 = \left(\frac{F_0 \tilde{\tau}}{2m\omega}\right) \sin \varphi$, $\frac{\delta p_0}{m\omega} = \left(\frac{F_0 \tilde{\tau}}{2m\omega}\right) \cos \varphi$. A subsequent ideal amplitude-and-phase measurement can reveal this change if the force $F_0$ exceeds the quantum limit [8]

$$F_0 \cong \left(\frac{2}{\tilde{\tau}}\right) (m\omega\hbar)^{1/2} \qquad (7).$$

No amplitude-and-phase measurement can do better than this.

The quantum limits in Eqs. 6 and 7 are traceable to the fact that $\hat{x}$ is not a continuous QND observable; a continuous measurement of $\hat{x}$ produces direct back action on $\hat{p}$, which then contaminates $\hat{x}$ through free evolution. On the other hand, $\hat{x}$ is a stroboscopic QND observable. Consequently, by stroboscopic measurements [5,9] at times $t = 0, \frac{\pi}{\omega}, \frac{2\pi}{\omega}, \ldots$ one can monitor $\hat{x}$ with perfect precision, in principle (except for the ridiculous limit from relativistic quantum theory, $\Delta x \geq \frac{\hbar}{mc} \cong 10^{-41}$ cm for m≅10 kg). stroboscopic measurements can be achieved with the system of Eq.1 by pulsing on and off the transducer's coupling constant K.

By a sequence of perfect stroboscopic measurements one can monitor an arbitrarily weak force $F_0$.

Perfect stroboscopic measurements require that $\hat{x}$ be coupled to the QRS for arbitrarily short time intervals $\tau$ at $t = 0, \frac{\pi}{\omega}, ...$ (and also that the QRS transfer its information to the first classical stage in a time less than $\frac{\pi}{\omega}$). If $\tau$ is finite then the momentum spread $\Delta p \geq \frac{\hbar}{2\Delta x}$, produced by a measurement of precision $\Delta x$, causes a mean position spread $\Delta x \cong (\frac{\Delta p}{m})\tau \geq \frac{\hbar \tau}{2m\Delta x}$ during the next measurement. The resulting r.m.s. error is [5,8,9]

$$\Delta x \geq (\frac{\hbar \tau}{m})^{½} \qquad (8).$$

The shorter the measurement time $\tau$, the more accurate the measurement can be.

Unfortunately, short measurements required very strong coupling of the antenna to the QRS in order to surmount the quantum mechanical zero-point energy that accompanies the signal through the QRS and into the amplifier.

If the phase $\varphi$ is near $\frac{\pi}{2}$ or $\frac{3\pi}{2}$, then the optimal times for the stroboscopic measurements are $t = 0, \frac{\pi}{\omega}, \frac{2\pi}{\omega}, ...$ and if the phase $\varphi$ is near 0 or $\pi$, then the optimal times for the stroboscopic measurements are obtained in ..., $t = \frac{\pi}{2\omega}, \frac{3\pi}{2\omega}$, Since the gravitational wave's phase is not predictable in advance, four antennas are required: a system consisting of two perpendicular antennas that are monitored in $t = 0, \frac{\pi}{\omega}, ...$ and another same system in $t = \frac{\pi}{2\omega}, \frac{3\pi}{2\omega}, ...$ [5,9].

**II. General model**

The Main difference in the structures of this system and of laser interferometer of gravitational wave detectors is that two perpendicular antennas are in-phase and according to figure1, we suppose their bisector coincide on x axis and y is the axis perpendicular to that. It is also supposed that gravity wave descends perpendicularly over the plate (x , y) and if two antennas have been located on x',y' axes which have gyrated toward x and y axes with the angle of 45º, then with the effect of gravity wave on (x´,y´) related to $h_+$ polarization which is one of the independent polarizations of gravity wave and with the assumption that we ignore the effects of $h_\times$ polarization as laser interferometer, then according to $h_+$ polarization of gravity wave, the antennas periodically contract and expand along x' and y'. Although gravity wave cannot be classically studied, its effect on the antenna makes classical investigating of its effect possible.

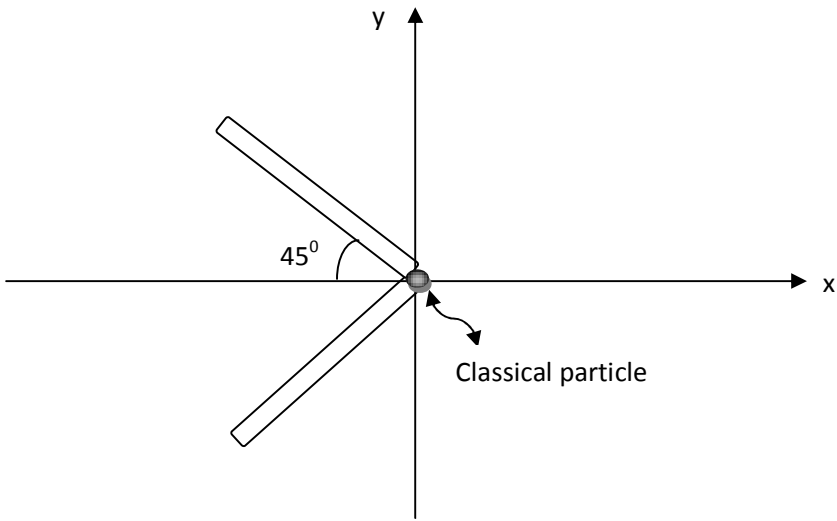

Figure1

So if we put a classical particle on the origin of coordinate of the above figure i.e. on the intersection point of the two antennas, then according to the gravity wave effect and displacement of antennas in each period and based on figure 2 we can conclude that based on momentum conservation law, momentum amount of $2p_y$ will be imposed on the particle along y axis at t=0 (the first period of gravity wave effect for phases φ= π/2, 3π/2, . . .). Since we have supposed that our two antennas are in-phase and in φ= π/2, 3π/2, . . ., then we do the measurements just in related stroboscopic times. So our measurements must be done in $t = 0, \frac{\pi}{\omega}, ...$ on a classical particle position.

In: $t = 0 , \varphi = \frac{\pi}{2}, \frac{3\pi}{2}, ...$

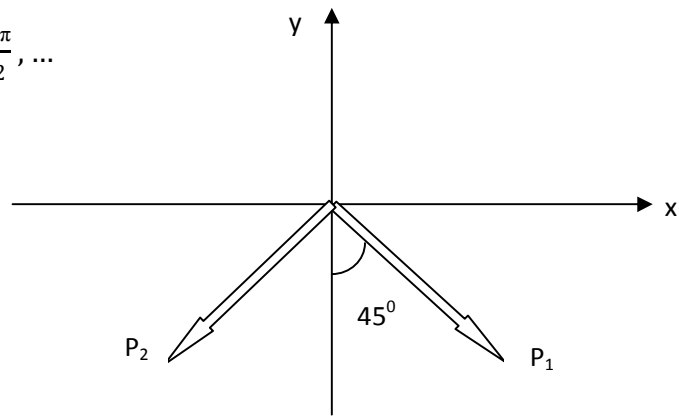

Figure2

We also need two force generators (a machine that can send a pulse or particle with a finite amount of momentum toward a classical particle in finite times) to put on y axis, one above the origin of coordinate and the other under the origin of coordinate with the same distance.

The system generally works in this way: in t=0, according to figure2 if the gravity wave is in the phase mentioned in this figure, then $p_x$s resultant of antennas becomes zero and $p_y$s resultant falls on the negative parts of y axis and according to figure3, the lower generator sends the pulse contemporarily in a way that we do our measurements only from classical particle position and with high accuracy. The same measurement is done for t=π/ω but at this time the difference is that the upper generator sends pulse and another classical particle is replaced in an interval between two measurements and measurement is done for it.

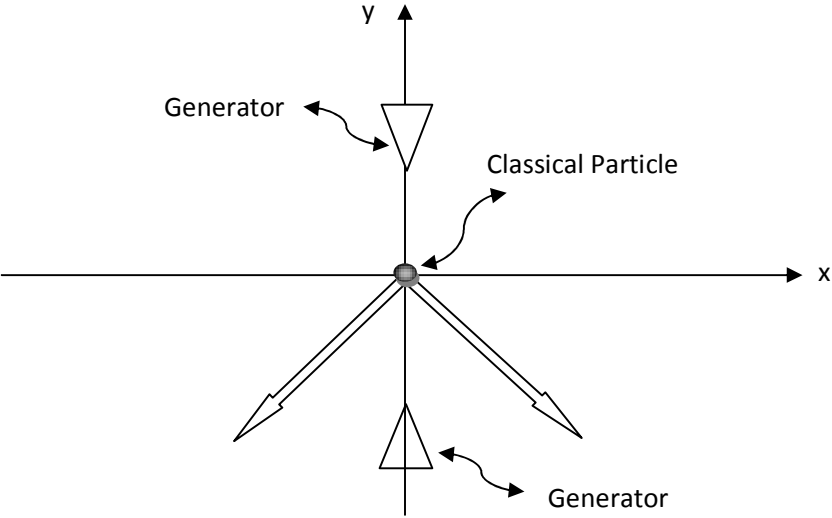

Figure3

Measurements will be repeated periodically so that each time pulse momentum amount of generator will change in the next measurement according to the direction (in the same or the opposite direction of y axis) and the amount of changes of the classical particle from its initial position after the momentum resultant effect of antennas and the sending pulse momentum on it, will change and will go in a direction that the particle displacement will reach to zero. When we would realize the particle has not been displaced with sending finite pulses, then sending pulse momentum will be equal to $2p_y$, and in this way, $p_y$ amount of each antenna will be obtained. Since p vector of antennas is in the angle of 45º with coordinate axes and because tan 45º= 1, So $p_x = p_y$ is a relationship that exists for both antennas. so we now have $p_x$ amount too and finally have both the amounts of $p_y$ and $p_x$ and according to this issue that the amount of sending pulses have high momentum accuracy, we can obtain momentum of gravity wave effect over each antenna with high accuracy. As we don't already know the phase of gravity wave, we simultaneously generate such a system in the same environment for phases $\varphi = 0, \pi, ...$ and stroboscopic times $t = \frac{\pi}{2\omega}, \frac{3\pi}{2\omega}, ...$ and do measurements as above in these times, until we cover all the possible phases too.

## III. Discussion

As mentioned above, although the gravity wave does not have classical features, as we study the effects of this wave only on the antenna and eventually on the classical particle, so we just contribute to precise measurement of the classical particle position which is a classical process and it is investigable.

The Another interesting point is that in this technical method, Heisenberg uncertainty principle does not impose any change in our measurement, since the classical particle and the antennas system are completely located outside of the generators system, such that there is no need to interaction and coupling between them and as a result it is adequate for us to have an accurate measurement for classical particle position in each stroboscopic time, without need to accurate measurement of its momentum. On the other hand we know that generators also send pulses by finite momentum and high accuracy in each time and only when we realized from position measurement of classical particle that its displacement is zero, above calculations will be completely done with study of pulse momentum being sent from related generator and at last we can calculate antenna momentum by this technical method with higher accuracy of standard quantum limit and minimum possible disturbance after gravity wave effect. Up to now no technical method has been invented to measure the momentum that has experimental aspect and we hope this method can solve the problem in spite of simplicity until at last by using it we will succeed in accurate measuring of gravity wave.

### Acknowledgment

We would like to thank Ali Rajaei for useful comments.